\documentclass[prd,nofootinbib,floatfix,twocolumn,amsfonts,showpacs]{revtex4}

\usepackage{graphicx,epsfig}
\usepackage{bm}
\usepackage{latexsym,amssymb,amsmath,float}

\newcommand{\be}{\begin{equation}}
\newcommand{\ee}{\end{equation}}
\newcommand{\bea}{\begin{eqnarray}}
\newcommand{\eea}{\end{eqnarray}}
\def\nn{\nonumber}

\def\trh{t_{RH}}

\def\arh{a_{RH}}

\def\ct{\eta}

\def\ctinf{\eta_{\infty}}
\def\cc{\Lambda}
\def\d{\Delta}
\def\ms{M_{\odot}}

\begin{document}

\title{Replacing Anthropy with entropy: Does it work?}

\author{Irit Maor$^1$,
        Thomas W.~Kephart$^{2}$,
        Lawrence M.~Krauss$^3$,
        Y.~Jack Ng$^4$ and
        Glenn D.~Starkman$^{1}$}
\affiliation{$^{1}$
    Center for Education and Research in Cosmology and Astrophysics,
    Department of Physics,
    Case Western Reserve University,
    Cleveland, OH 44106-7079, USA}
\affiliation{$^{2}$
    Department of Physics and Astronomy,
    Vanderbilt University, Nashville,
    TN 37235, USA}
\affiliation{$^{3}$
    School of Earth and Space Exploration,
    and Department of Physics,
    Arizona State University,
    Tempe, AZ, 85287-1404, USA}
\affiliation{$^{4}$Department of Physics and Astronomy,
    University of North Carolina,
    Chapel Hill, NC 27599-3255, USA}

\begin{abstract}
Probably not, because there are lots of manifestly unanthropic ways of
producing entropy.  We demonstrate that the Causal Entropic Principle (CEP),
as a replacement for the anthropic principle to explain the properties of the observed
universe, suffers from many of the same problems of
adopting myopic assumptions in order to predict that
various fundamental parameters take approximately the observed values.
In particular, we demonstrate that four mechanisms --
black hole production, black hole  decay,  phase transitions, and dark
matter annihilations or decays -- will manifestly change
the conclusions of the CEP to predict that we should live in a
universe quite different than the one in which we find ourselves.
\end{abstract}

%\begin{keywords}
%cosmology
%\end{keywords}

\pacs{99}

\maketitle

\section{Introduction}

Recently, it was proposed \cite{Bousso:2007kq} that the correct anthropic measure for the probability of observing a given universe is tied to the amount of entropy production in that universe. According to this proposal, known as the Causal Entropic Principle (CEP),
life depends on the production of entropy and we are more likely to live in a universe
in which the entropy production is as large as possible. The entropy should be calculated within a causally connected 4-volume known as the causal diamond \cite{Bousso:2006ev}. The authors of \cite{Bousso:2007kq} determined that the most important sites of  entropy production in our universe are the radiating dust grains heated by stars. Life should therefore cluster around stars -- a postdiction consistent with the lone observed nexus of living organisms -- and the parameters of our universe should maximize the number density of stars and associated dust clouds. Keeping all other parameters fixed, the value of the cosmological constant that maximizes the dust entropy within the causal diamond is in reasonable agreement with the observed low but non-zero value.

However, it is not clear that interstellar dust is indeed required for life, e.g., if we strip off all the dust from the solar system then the entropy production would be reduced but life on earth would be virtually unaffected. Life does seem to need a steady gentle (over several billion years) source of free energy, which intuitively is associated with semi-adiabatic processes that produce little entropy. Violent energy release and concomitant entropy production is unlikely to be conducive to the formation of life which
depends on complicated fragile molecules.

Furthermore, if we do assume ourselves to be typical representatives of observers, it seems that baryons are essential to the formation of life. The CEP does not accommodate this fact.
The baryon to entropy ratio (B/S) can be very small, (e.g., in our universe B/S$\sim 10^{-9}$) or even vanishingly small if the Sakarov conditions  (baryon number violation, T violation and CP violation) are not satisfied. Such a baryon poor universe is certainly not a hospitable environment for life. Considering the ~$10^{500}$ possible vacua in the string landscape, there are surely many examples of universes with multiple first order phase transitions and lots of entropy production, but without enough baryons for nucleosynthesis to proceed and no life. Such large entropy producing universes would be a detriment to the formation of life. By assuming ourselves to be typical representatives of observers we essentially limit the above discussion to carbon-based forms of life similar to those we find on earth. We are adopting here a conservative approach: relaxing the assumption of carbon-based life opens up an additional volume of parameter space where life can develop and the entropy measure is simply unknown.

Three sources of entropy which were either not accounted for or were discounted in \cite{Bousso:2007kq} are the entropy associated with black holes, the entropy produced at reheating after a phase transition, and the entropy produced by out of equilibrium decays or annihilations of dark matter particles, presumably because none seems to be directly related to life. One could argue that as long as the entropy of the black hole is behind the horizon it is irrelevant to life and to the CEP. However, if there is a way to ``mine" this entropy, or if it is radiated out of the black hole, then in principle this is no different than other sources of entropy. As we explicitly show below, this source of entropy is by far the most significant source, and changes the CEP predicted value of the cosmological constant. The entropy of phase transitions, reheating, decays and annihilations is what is referred to in \cite{Bousso:2007kq} as ``matter sector entropy", which is not different in nature than say the heated dust entropy.
We present a calculation showing that if we optimize the entropy production by reheating
then inflation will not result in the required number of e-folds of growth of the universe.
We also show that dark matter annihilation can lead to far greater
entropy production than currently results from stellar burning.

The paper is organized as follows, the entropy of black holes and its accessibility are considered in section \ref{bh}. In section \ref{inf} we maximize entropy produced after a phase transition, and show it is inconsistent with the observed universe. Out of equilibrium processes are discussed in section \ref{ooe}, and we conclude in section \ref{disc}.

\section{Entropy and black holes}
\label{bh}

It has been argued \cite{Penrose, Kephart:2002bf} that most of the entropy of our universe is in the form of the supermassive black holes (SMBH) found at the centers of galaxies.
Indeed, the Hawking-Bekenstein entropy,
\bea
  S_{HB} = 4\pi G_N M^2  %_{\rm bh}
  \label{HBentropy}
\eea
of one black hole of mass  $M\simeq 10^8 \ms$ is $S_{HB} \simeq 10^{93}$, approximately 20000 times the entropy of the CMB within the entire observable universe -- the next greatest observed reservoir of entropy --
\begin{eqnarray}
  \label{SCMB}
  S_{CMB} &\simeq& \frac{2970}{{\rm cm}^{3}}\left(\frac{{\rm T}}{2.75{\rm K}}\right)^3
        \frac{4\pi}{3} \left(\int_0^{t_0}\frac{a_0}{a(t')}dt'\right)^3\\
  &\simeq& 5\times 10^{88} \nonumber
\end{eqnarray}
(for $H_0=72$km/s/Mpc).

Given that the energy density of matter (dark and baryonic) today is about $0.3\times 10^{-48}$ GeV$^4$, consolidating the mass of the whole universe to a single black hole would give entropy of
\begin{eqnarray}
  && S(M_{universe})\simeq 8\times 10^{120}
\end{eqnarray}
which exceeds by far the CMB entropy (it is assumed here that dark and baryonic matter have similar entropic properties).  As a matter of fact, the mass of the universe can be divided into about $10^{29}$ black holes and still contain more entropy than the CMB. The diamond cannot shrink onto a black hole before the black hole evaporates because the metric in its immediate vicinity will not be de-Sitter like. However, even without considering such extreme scenarios, we can compare the entropy of star-heated dust and black holes in the observed universe. Since at least 30\% of the spiral and disc galaxies that have been checked \cite{centralblackholes} have been found to contain a black hole of mass between approximately
$3 \times 10^6$ and $ 3\times10^9 M_\odot$, and since there are approximately $10^{11}$ such galaxies within the observable universe, SMBHs probably contain $10^{13-15}$ times more entropy than all other reservoirs combined.

The process of accretion onto a black hole is an ongoing one. The black hole at the center of our galaxy is being fed by infalling matter at a rate of $\approx$few$\times 10^{-6}M_\odot$/yr \cite{Cuadra:2005pt}.
Differentiating \eqref{HBentropy}, we see that the rate of entropy production
due to accretion onto a black hole is:
\begin{equation}
  \label{HBentropydot}
  \dot{S}_{HB} =  8\pi G M \dot{M}.
\end{equation}
The entropy production in the universe due to black hole accretion is thus
\begin{equation}
  \label{cosmic_entropydot}
  \dot{S}_{cosmic} =  8\pi G  \sum_{galaxies}  M \dot{M}.
\end{equation}
Taking $\dot{M}$ to be about the Milky Way value, adopting an approximate value of $10^8M_{\odot}$ for $M$, and taking there to be approximately $10^{11}$ galaxies in the observable universe, we find
\begin{equation}
  \label{cosmic_entropydot_b}
  \dot{S}_{cosmic} \simeq 10^{90} /{\rm yr}.
\end{equation}
This is vastly larger than the entropy production by any other known source. In particular, the authors of \cite{Bousso:2007kq} find that the infrared emission from dust-grains surrounding stars produces an entropy of  $\sim\!\!10^6$ per baryon over the life-time of the universe.
However, the dust grain entropy is only about one part in $10^3$ of the CMB entropy. Comparison between light element abundances and the predictions of standard big bang nucleosynthesis yield a baryon-to-photon ratio of $\eta_{BBN}\simeq5\times10^{-10}$.
Given the value of $S_{CMB}$ found above,
\begin{equation}
\dot{S}_{stellar} \simeq few \times 10^{75}/{\rm yr} .
\end{equation}

One may perhaps object that  the Hawking-Bekenstein entropy produced by the growth of a black hole is not actually produced (or at least not made accessible) until the black holes decay by way of Hawking radiation.  The larger the mass of the black hole, the longer this takes, and
certainly for the black holes at the centers of galaxies this time scale is considerably longer than the age of the universe.  This claim would seem to fly in the face of the usual connection between black-hole area and entropy required by the second law of thermodynamics.
Nonetheless, even that will not prevent black holes from being the primary source of entropy
in the universe.

It is not necessary to wait for a black hole to decay naturally. Unruh and Wald showed \cite{UnruhWald} that the energy of a black hole can be  mined at a rate limited by $\left(\frac{dE}{dt}\right)_{UW} \sim 1/D^2$ (where $D$ is the proper distance from the horizon), 
\begin{equation}
  \left(\frac{dE}{dt} \right)_{max} \simeq m_{Pl}^2 \simeq 10^{79} ~{\rm eV/ yr}.
\end{equation}
which is far above the natural Hawking emission rate.
Given the entropy of the black hole (\ref{HBentropy}), this amounts to an entropy release
rate of
\begin{equation}
  \dot S_{UW} \simeq 8\pi M \simeq 10^{90} \frac{M}{M_\odot} {\rm yr}^{-1} .
\end{equation}
With approximately $10^{11}$ black holes of mass $10^8M_\odot$ in the observable universe,
the total entropy release rate could reach $10^{109}{\rm /yr}$, far above the rate due to the emissions of interstellar dust. A word of caution should be added regarding this method of mining black holes, as it seems in danger of violating the second law of thermodynamics, because too little entropy is extracted per unit of energy by mining to add up to A/4.

While this method of extracting entropy from a black hole seems to require active intervention
by intelligent observers, another process happens inevitably. The process of accretion onto a black hole is itself not particularly efficient. Simulations suggest \cite{energyloss} that, in mergers of comparable-mass black-holes, typically ${\cal O}(10\%)$ of the total mass of the system emerges as gravitational radiation, with a wavelength comparable to the Schwarzschild radius of the black holes, $r_S=2GM$. Given that most galaxies seem to have central black holes, and that most galaxies  seem to have undergone at least one major merger, we can estimate that
 \begin{eqnarray}
 \label{Smergerdot}
 \dot{S}_{\rm mergers} &\simeq&  N_{\rm galaxies} \frac{0.1 M_{\rm central-bh}}{2\pi/r_S}
			\frac{1}{t_0} \nonumber \\
			&=& 0.1 N_{\rm galaxies} \frac{M_{\rm central-bh}^2}{m_{Pl}^2 t_0}\\
			&\simeq& 10^{90}/{\rm yr} \nonumber
 \end{eqnarray}
Again, this vastly exceeds the entropy production rate due to stellar irradiation of dust.

For the CEP to be consistent with the observation that most of the entropy production in the universe is tied up with black hole accretion, we would require  that life is somehow closely associated with accretion onto black holes. This connection is not immediately clear to us.  Neither the accretion of matter onto central black holes, nor the emission of a large flux of very long-wavelength gravitational radiation seems particularly essential to the emergence of life as we know it. Moreover, if entropy is a good proxy for observations and observers, one would expect life to emerge close to entropy sources even within our universe.
We are far out in the galaxy, i.e., far away from the  entropy producing source.
One might expect that we would live nearer the black hole.
Our own galaxy's central black hole is a meager few $10^6$ solar masses, whereas the largest observed black holes are well above $10^9$. By the conventional anthropic reasoning applied to \eqref{Smergerdot}, we would have been very unlikely ($\sim10^{2(6-9)}=10^{-6}$) to find ourselves in a galaxy with such a small central black hole.

At any rate, it does not seem to be the case that we live in a universe optimized for the
production of entropy. Inspection of \eqref{Smergerdot} suggests that  such a universe would
have as much as possible of the energy density concentrated in a much smaller number
of very large black holes.  These black holes would be fed by a maximized inflow of accreting matter.

\section{Entropy and inflation}
\label{inf}

While phase transitions typically produce less entropy than supermassive BH horizons, they still produce substantially more than stellar heated dust. Reheating (see \cite{Kofman:1994rk}, \cite{Kofman:1997yn} and \cite{Shtanov:1994ce} for details) produces copious amounts of entropy. Parametric resonance leads to a reheat temperature higher than would otherwise be expected (i.e., $T > 10^9 $ GeV), and hence allows for multiple phase transitions before reaching a low temperature, low $\Lambda$ state. A causal diamond can therefore contain multiple reheating surfaces.

Consider a (flat) universe dominated by vacuum energy density $\rho_I$, the inflaton. During this period the scale factor grows exponentially, $a(t)\sim\exp[t]$. At time $\trh$, this early vacuum energy density decays to matter, with the assumption that the process is instantaneous and $100\%$ efficient. Ignoring the fine details of reheating, the amount of entropy density produced is approximately $\exp[3N]\rho_I^{3/4}$, where $N$ is the number of e-folds the universe expanded. The scale factor evolution then slows down to the standard matter dominated behavior, $a(t)\sim t^{2/3}$. We also assume that there is some minute residual vacuum energy density $\rho_\cc$ which did not decay, and will eventually overcome the matter energy density to introduce a late time inflationary phase. We recognize $\rho_\cc$ as the cosmological constant. It is convenient to define the ratio $R = \rho_I/\rho_\cc$, and we expect this number to be very large.

The essential elements of this set up are the early inflation as a source of entropy, and the late inflation. The latter is necessary to make sure the universe becomes de-Sitter at old age, so the total conformal life time is finite. Without this late time de-Sitter period the definition of a causal diamond is ambiguous{\footnote{The ambiguity in the definition of the causal diamond prevents consideration of universes where conformal time is not finite. This raises questions as to the validity of the causal diamond being used in such a statistical manner.}}.
With this description of the universe, we are ignoring the radiation-dominated phase between inflation and matter domination. As shown in the appendix of \cite{Bousso:2007kq}, this is a reasonable model of our universe, and ignoring the radiation era introduces only a negligible error. It also has the advantage of simplifying consequent expressions.

Unlike \cite{Bousso:2007kq} though, we are {\it not} going to ignore the reheat entropy produced by the decay of the inflaton. Accordingly, we will tune the edge of the causal diamond (the time which the 3-volume of the diamond is maximal) to the time of entropy production due to the inflaton decay, $\trh$, as to optimize the entropy production within the diamond. Thus we demand that
\bea
 \eta_{edge}=\eta_{RH}
\eea
where $\eta=\int dt/a(t)$ is conformal time. In order to calculate $\eta_{edge}$, we define the generating point of the forward cone of the causal diamond at $\ct(t=t_i)\equiv 0$ and the backward cone at $\ctinf=\ct(t=\infty)$. The edge of the diamond, where the two cones intersect, occurs at
\bea
  \eta_{edge}=\frac{1}{2}\ctinf &=&
      \frac{1}{2a_iH_i}
         \Bigg(\frac{9}{2}(N+1)\exp[-N]R^{1/6} \nn \\
         &+& 1-(N+2)\exp[-N]
         \Bigg)
\label{edge}
\eea
where $a_i$ and $H_i$ are the values of the scale factor and the Hubble function at the initial point of the causal diamond. On the other hand, a simple calculation gives
\bea
  \eta_{RH}=\frac{1}{a_iH_i}\left(1-\exp[-N] \right)
\label{RH} ~.
\eea
Combining \eqref{edge} and \eqref{RH} gives
\bea
  R = \left(\frac{2}{9}\left(\frac{\exp[N]+N}{N+1} \right) \right)^6
     \approx \frac{\exp[6N]}{(4.5N)^6} ~,
  \label{rmd}
\eea
where the approximation is for large values of $N$.

Equation \eqref{rmd} relates between the number of e-folds produced during inflation, the energy density of the inflaton and the value of the present cosmological constant. It is a direct result of synchronizing between the edge of the causal diamond and the time of reheating. Thus any results derived from this equations should be interpreted as a {\it consequence} of the CEP. We now show two such erroneous consequences. \\

{\bf Entropy production}

Ignoring fine details of reheating and assuming the process to be instantaneous and completely efficient, the entropy produced as the inflaton decays is
\bea
 && \d S \propto \left( \arh T_{RH} \right)^3 =\exp[3N]\rho_I^{3/4}
 \label{ds}
\eea
where we took the reheat temperature to be $T_{RH} = \rho_I^{1/4}$. If indeed the entropy production in a diamond is the statistical weight of that vacuum, then we want to maximize $\d S$. The more $N$ and $\rho_I$ the merrier:
$N=N_{*}$ and $\rho_I=\rho_{*}$, where the starred values stand for the maximal allowed value. For the inflaton energy density, this is presumably $m_{Pl}^4$. Applying these values to eq.~\eqref{rmd} we can estimate what value of $\rho_{\cc}$ is preferred by entropic considerations:
\bea
 && \rho_{\cc}=\frac{\rho_I}{R}=\rho_{*}(4.5N_{*})^6\exp\left[-6N_{*} \right] \rightarrow 0
\eea
The value of the cosmological constant is exponentially suppressed with the number of e-folds. In this scenario, entropic considerations predict a zero cosmological constant. \\

{\bf Number of e-folds}

\begin{figure}
  \begin{center}
    \epsfig{file=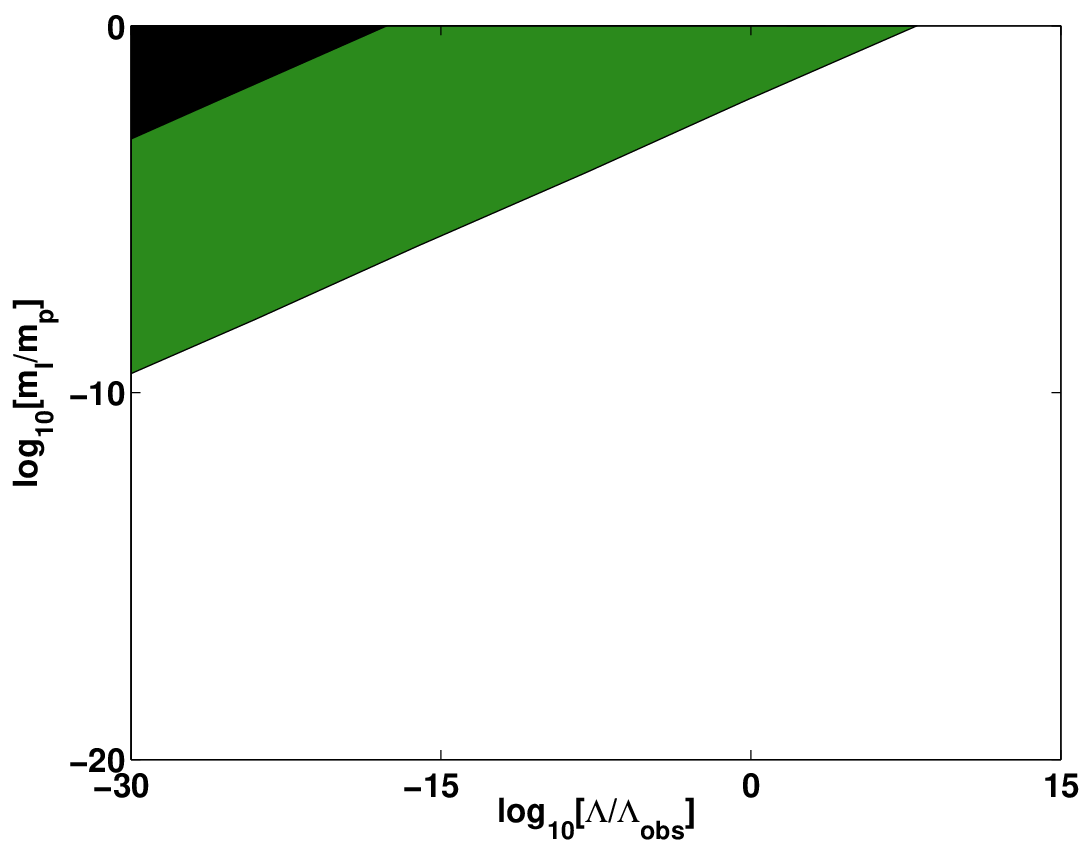,height=60mm}
  \end{center}
%  \vspace{-0.6cm}
  \caption{A contour plot of the number of e-folds produced during inflation as a function of the cosmological constant (normalized by the observed value, $\Lambda_{obs}=0.7\times 10^{-48}$ GeV$^4$ and the inflaton mass (normalized with the plank mass). The relation plotted assumes the CEP holds, and tunes the edge of the causal diamond to the time of reheating. The shaded area indicates where $N>60$ (black) and $N>50$ (green).
  \label{efolds}}
\end{figure}

Another interesting exercise is to see whether eq.~\eqref{rmd} is consistent with our knowledge of our universe. We know that inflation needs to produce a minimum of 60 e-folds in order to solve the classical problems of flatness, isotropy, etc. The exact number of e-folds needed depends on the details of the inflationary model and in particular on the reheating process \cite{Podolsky:2005bw}, and in some cases can go lower than the canonical value of 60.  However, for our simplistic order of magnitude estimate we assume that reheating is instantaneous and 100\% efficient, and we will be conservative by demanding at least 50 or so e-folds. We can ask ourselves what values of $\rho_\cc$ and $\rho_I$ provide enough e-folds, under the constraint of \eqref{rmd}. Figure \ref{efolds} shows contours of $N=60$ and $50$ over the parameter space of the inflaton energy scale $m_I/m_{Pl}$ and the energy density of the cosmological constant $\Lambda/\Lambda_{obs}$ (where we took $\Lambda_{obs}=0.7\times 10^{-48}$ GeV$^{4}$). In order to produce enough e-folds, the mass of the inflaton needs to be exceptionally high, together with an exceptionally low mass for the cosmological constant. For example, taking acceptable values of $\cc=10^{-48}$ GeV$^4$ and $m_I=10^{15}$ GeV produces only 46.7 e-folds, which is insufficient.

%%%%%%%%%%%%%%%%%%%%%%%%%%%%%%%%%%%%%%%%%%%%%%%%%%%%%%%%%%%%%%%%%%%%%
%%%%%%%%%%%%%%%%%%%%%%%%%%%%%%%%%%%%%%%%%%%%%%%%%%%%%%%%%%%%%%%%%%%%

\section{Out of Equilibrium Dark Matter Annihilation or Decay}
\label{ooe}
Over the course of its history, particle-antiparticle annihilations as mass thresholds have been crossed has dumped significant entropy into what is now the radiation bath. These annihilations, however, have occurred close to  thermal equilibrium, so that total entropy was nearly conserved, and merely transferred from one relativistic species to another adiabatically.

Significant entropy can, however, be generated by the eventual disappearance of the massive weakly interacting particle species which are presumed to dominate the energy density of large scale  structures in the Universe in the form of dark matter.  The remnant number density of WIMP dark matter is determined by the point at which thermal annihilations in the early universe fall out of equilibrium as their number density decreases to the point where their annihilation rate falls below the expansion rate of the Universe.   However, once a structure forms its dark matter density remains constant.  After a sufficiently long time, then, the age of the universe will exceed either the lifetime of the dark matter particle (if it is unstable) or the mean free time per particle for annihilation.   The dark matter within gravitationally bound structures will then completely decay or annihilate away.

Annihilation or decay alone, even out of equilibrium, do not necessarily result in significant entropy generation. If the annihilation products are equal in number to the dark matter particles, or only outnumber them by a small factor, then the total entropy produced will not approach that emitted by stellar burning, or, as \cite{Bousso:2007kq} argues, by dust warmed by starlight.   However, just as the entropy production from nuclear burning is greatly enhanced by the multiple number changing interactions of the decay products, so too is the case for dark matter. Indeed, there are two ways that this enhancement takes place.

Although the dark matter particles presumably decay or annihilate directly into just a very few standard model particles, the final decay products can only be the stable standard model particles: electrons, positrons, photons, neutrinos, antineutrinos, and protons.   While neutrinos and antineutrinos will stream freely out of almost any large scale structure, not so for the rest --
they are more likely to have interactions.  For the photons, the probability of interaction depends on their frequency -- the lower the better -- and on the location from which they were emitted.  Photons emitted in optically dense regions may well thermalize and result in considerable entropy production.  But let us focus instead on the charged particles, in particular the electrons.

There are several ways in which high energy electron can lose some energy and in the process emit
a very large number of low-energy photons. One of those is synchrotron radiation in the presence of a magnetic field. A 10GeV electron, such as might be a decay or annihilation product of a 100GeV dark matter particle, traveling for $10^6$ years (about how long it takes to cross a cluster, and so an underestimate since cosmic ray electrons can wander in the local magnetic field) in the presence of a one micro-Gauss magnetic field (typical of clusters) will emit about  $10^{12}$ synchrotron photons.   Given that the typical cluster contains about $3\times10^{70}$ dark matter particles (at a mass of $100$GeV), the annihilation of cluster dark matter within one cluster will result in an entropy production of at least
\be
\Delta S_{\rm DMsynch} \simeq 3\times 10^{82} .
\ee
This is approximately the equivalent of $10^{7}$years of stellar production within the entire current de Sitter volume. This means that if dark matter decay or annihilation were occurring now, when there are about $10^6$ clusters within the causal diamond, then entropy production resulting
from dark matter would significantly exceed anything due to stars.  Alternately, if the cosmological constant were considerably lower then the epoch of  dark matter annihilation {\em would} occur near the time when the three-volume of the  causal diamond  was maximal and associated entropy production would exceed that due to stellar burning.

The  CEP would seem to suggest that we should be live in a universe that better utilizes its dark matter to generate entropy, and in closer association with this entropy production, either temporally or spatially. One can easily imagine this source of entropy being far larger than is expected in our universe. A decrease in the cosmological constant, an increase in the dark matter density (relative to photons), or an increase in the amplitude of fluctuations (so that larger structures can form before the onset of cosmological-constant-domination) would all increase the number of dark matter annihilations within the largest bound structures.   Meanwhile, an increase in  the magnetic field inside the largest bound structures would result in more synchrotron radiation. Note that synchrotron radiation is not the only possible mechanism of efficient entropy generation.  Additional possible scattering processes off of matter, including bremsstrahlung emission, could in some cases, also be significant.

\section{Discussion}
\label{disc}
In our universe, life does not seem to be associated with either the times or locations of
the greatest entropy or the greatest entropy production -- formation of, or accretion onto,
massive black holes at the centers of galaxies, post-inflationary (p)reheating, or synchrotron radiation from  charged particles produced in late-time dark matter annihilation and decay.

Furthermore, we can look at our own universe and ask how we could increase the rate of entropy production. A higher  dark matter density and higher amplitude of primordial fluctuations
would have resulted in the formation of denser  more massive galaxies and clusters.
Energy loss by dark matter annihilation (or decay) products in those structures would
have resulted in greater entropy production.
Denser galaxies would also have a higher rate of accretion onto their central black hole,
resulting in both larger black holes, and more entropy released into the environment during the accretion. These denser galaxies would have occupied richer, denser clusters  resulting in more galaxy mergers, and thus higher final black hole masses.    More massive black holes implies a corresponding increase in the black hole entropy -- entropy which is ultimately released through Hawking radiation of ordinary photons (and other standard model particles) into a single causal diamond. These denser galaxies and clusters could also have accommodated a
higher value of the cosmological constant than we currently see without being disrupted.

Meanwhile, if we keep the dark matter density and the amplitude of primordial fluctuations fixed at their values in our universe,  a lower value of the cosmological constant would have allowed the formation  of larger gravitationally bound systems -- super clusters, super-duper-clusters and so-on.  These more massive bound systems would eventually  have had more massive black holes at their centers, eventually resulting in more entropy release within a single causal diamond than in our universe.

Thus, life does not seem to have demanded that the universe optimize the production of
entropy.  Most of the entropy seems not to be particularly closely associated with the assumed
concentration of life around  stars. A modification of the CEP to count only standard model entropy
could no doubt be formulated, but would not evade the problem that dark matter annihilations into
standard model particles results in enormous entropy production via synchrotron radiation.
Moreover,  as with much anthropic reasoning it would seem to be another example of designing
the questions to give the already known answers.

\section*{Acknowledgments}

The work of TWK was supported by U.S. Department of Energy grant number  DE-FG05-85ER40226.
IM is supported by DOE.
YJN is supported in part by DOE.
GDS is supported by NASA ATP and DOE grants to the CWRU
particle-astrophysics theory group.
GDS also thanks Maplesoft for use of the \textsc{Maple} software.

\end{document}